\documentclass[aps,pre,superscriptaddress,preprint]{revtex4-1}
\usepackage[latin1]{inputenc}
\usepackage{amsmath}
\usepackage{amsfonts}
\usepackage{bbold}
\usepackage[b]{esvect}
\usepackage{color}
\usepackage{graphicx}
\usepackage{hyperref}
\hypersetup{linktoc=page}
\hypersetup{
    colorlinks,
    citecolor=blue,%    filecolor=black,%
    linkcolor=blue%     urlcolor=black
}

\makeatletter
\newcommand{\DeclareMathMacro}[2]{%
  \expandafter\let\csname original@\expandafter\@gobble\string#1\endcsname=#1
  \expandafter\def\csname\expandafter\@gobble\string#1\endcsname{\relax\ifmmode#2\else\csname original@\expandafter\@gobble\string#1\expandafter\endcsname\fi}
}
\makeatother

\DeclareMathMacro{\e}{\mathrm{e}} % e tel que ln(e)=1
\DeclareMathMacro{\i}{\mathrm{i}} % i tel que i^2=-1
\DeclareMathMacro{\R}{\text{\textbb{R}}}

\renewcommand{\Re}{\mathop{\mathrm{Re}}}
\renewcommand{\Im}{\mathop{\mathrm{Im}}}

\newcommand{\diff}{\mathop{}\mathopen{}\mathrm{d}}
\newcommand{\dt}{\diff t}

\begin{document}
\title{Stimulated Raman scattering in the relativistic regime  in near-critical plasmas}
\author{J. G. Moreau}
\affiliation{Universit\'e de Bordeaux-CNRS-CEA, Centre Lasers Intenses et Applications, UMR 5107, F-33405 Talence, France}
\author{E. d'Humi\`eres}
\affiliation{Universit\'e de Bordeaux-CNRS-CEA, Centre Lasers Intenses et Applications, UMR 5107, F-33405 Talence, France}
\author{R. Nuter}
\affiliation{Universit\'e de Bordeaux-CNRS-CEA, Centre Lasers Intenses et Applications, UMR 5107, F-33405 Talence, France}
\author{V. T. Tikhonchuk}
\affiliation{Universit\'e de Bordeaux-CNRS-CEA, Centre Lasers Intenses et Applications, UMR 5107, F-33405 Talence, France}
\date{\today}
\begin{abstract}
Interaction of a high intensity short laser pulse with near-critical plasmas allows to achieve extremely high coupling efficiency and transfer laser energy to energetic ions. One dimensional Particle-In-Cell (PIC) simulations are considered to detail the processes involved in the energy transfer. A confrontation of the numerical results with the theory highlights a key role played by the process of stimulated Raman scattering in the relativistic regime. The interaction of a 1 ps laser pulse ($I \sim 6 \cdot 10^{18}$ W.cm$^{-2}$) with an under-critical ($0.5\, n_c$) homogeneous plasma leads to a very high plasma absorption reaching 68 \% of the laser pulse energy. This permits a homogeneous electron heating all along the plasma and an efficient ion acceleration at the plasma edges and in cavities.
\end{abstract}
%\pacs{PACS: 52.50.Jm, 52.75.Di, 25.20.-x}
\maketitle

\section{Introduction}
\label{sec_intro}

Ion acceleration with intense laser pulses is promising for applications in radiography, inertial confinement fusion and radiotherapy \cite{borghesi2002, roth2001, malka2004}. The well known methods of ion acceleration - target normal sheath acceleration (TNSA) \cite{hatchett2000} and radiation pressure acceleration (RPA) \cite{esirkepov2004, macchi2005} - consider thin solid targets, which are not transparent for the incident laser radiation. This approach implies a controlled laser pulse temporal contrast and is not suited for applications because of a relatively low coupling efficiency, difficulties to refresh targets in high repetition rate regime and high energy projectiles and debris. Use of low density targets offers an attractive alternative as the interaction takes place in a plasma volume and targets could be refreshed more easily with much less debris. One version of this approach is the break-out afterburner (BOA) scheme \cite{yin2007} where the initially solid target becomes transparent during the interaction so that the initially TNSA-accelerated ions are further re-accelerated in the expanding target plasma. 

Another approach consists in using foams \cite{willingale2009}, gas jets \cite{matsukado2003,dhumieres2010,dhumieres2013PoP} or exploded foils \cite{Antici2009,dhumieres2013PPCF} with plasma densities smaller or comparable with the laser critical density. Laser pulses can penetrate through such targets allowing a more efficient transfer of energy to electrons and ions. Propagation of intense laser pulses in near critical plasmas is characterized by the effect of  relativistic transparency \cite{akhiezer1956, kaw1970}. It implies that a plasma with the overcritical electron density $n_e> n_c$ can be transparent for the laser wave if its intensity satisfies the condition \cite{kaw1970}
\begin{equation}\label{eq1}
n_e< n_c(1+a_0^2/2)^{1/2} .
\end{equation} 
Here, $a_0=eE_0/m_e \omega_0 c$ is the dimensionless laser amplitude, $n_c=m_e\varepsilon_0 \omega_0^2/e^2$ is the critical electron density, $m_e$ and $-e$ are the electron mass and charge, $c$ is the velocity of light in vacuum, $\varepsilon_0$ is the dielectric permittivity of vacuum and $\omega_0$ is the laser frequency. This condition is written for a linearly polarized wave and does not account for other processes that may take place at relativistic laser intensities, $a_0>1$, such as relativistic self-focusing \cite{max1972}, parametric instabilities  or density profile steepening by the laser ponderomotive force \cite{guerin1996, li2008, iwawaki2015}. The parametric instabilities \cite{sakharov1994, guerin1995} and electron acoustic modes \cite{fernandez2000, vu2001, nikolic2002} excited by the intense laser pulse may significantly perturb its propagation. 

Although the experiments show promising results concerning ion acceleration in near critical plasmas \cite{gauthier2014}, there is no clear understanding of the mechanisms of laser energy transfer and ion acceleration. The condition \eqref{eq1} is necessary but not a sufficient condition for the relativistic plasma transparency. The leading edge of the pulse exercises a ponderomotive force  creating the electron pileup, which may lead to a partial or complete laser reflection. Such regimes of the plasma piston supported by the laser radiation pressure were considered in Refs. \cite{guerin1996, iwawaki2015}. They correspond to a relatively slow laser penetration into the plasma and a less efficient energy coupling. A red shift of the backscattered light is considered as a signature of the Doppler shift of the laser light reflected from the moving piston. In contrast, for a sufficiently low plasma density, $n_e<n_{\rm th}$, the laser pulse can propagate without strong reflection. The threshold density for a circularly polarized wave reads \cite{cattani2000, schep2000, siminos2012}:
\begin{equation}\label{eq2}
n_{\rm th}= \frac{1}{2}n_c \left(1+\sqrt{1+2a_0^2}\right) \text{ in the limit } n_{\rm th}<3/2\, n_c
\end{equation} 
and it scales as $n_{\rm th} \propto a_0$ for large amplitudes $a_0\gg1$. This regime of near critical plasma density  $n_e \lesssim n_{\rm th}$ seems to be the most appropriate for efficient laser energy coupling to plasma.  

In this paper, we present a detailed study of the interaction of short intense laser pulses with a near critical plasma for mildly relativistic conditions ($a_0\sim 2$). By using Particle-In-Cell (PIC) simulations, we show that this regime leads to a very efficient energy transfer to electrons via the process of Stimulated Raman Scattering (SRS) in the relativistic regime. This energy is then transmitted from electrons to ions.

This paper is organized as follows. In Sec. \ref{sec_mainresults}, we present our main PIC simulations results. After describing the simulation parameters, we study the absorption of the laser pulse by the plasma and the propagation of the electromagnetic waves in it, in Sec. \ref{sec_plasmaabsorption}. We then present the time-frequency analysis of the electromagnetic and electrostatic waves in the plasma and vacuum, in Sec. \ref{sec_spectralaanalysis}. This analysis demonstrates a development of the SRS instability. This observation is further confirmed in Sec. \ref{sec_SRSconfirmation} by the wave vector analysis and comparison with an analytical model for the relativistic laser pulse in a cold plasma \cite{guerin1995}. We then analyse the electron heating and the ion acceleration in Sec. \ref{sec_electronheating}. Finally, we discuss these results by comparisons with other numerical simulations and give our conclusions in Sec. \ref{sec_discussion}.

\section{Particle-In-Cell simulation results}
\label{sec_mainresults}

The numerical simulations are performed with the fully electromagnetic relativistic PIC code OCEAN \cite{nuter2013} in the 1D3V geometry. 

The numerical noise was strongly suppressed by using a third order interpolation function for the macroparticles. The mesh length $\Delta x=0.00796\,\lambda_0$, the time step $\Delta t=0.00796\,T_0$ and the number of macroparticles per mesh $N_{mpm}=750$ were chosen so that the numerical heating of the macroparticles during the calculation was maintained at a level lower than 0.07\% of the laser energy. We measured the electron and ion energies in the plasma, the instantaneous and cumulated reflectivity and transmission, and the electrostatic and electromagnetic energies in the simulation box. The collisions are not accounted for in these simulations as the characteristic collision time is longer that the run time and the dominant physical processes are related to the parametric instabilities and wave-particle interactions. 

In what follows, the lengths and times are normalized to the laser wavelength $\lambda_0=1\,\mu$m and the period $T_0=\lambda_0/c \approx 3.3$ fs, respectively. The electric and magnetic fields are normalized to the Compton fields $E_c = m_e c\omega_0/e \approx 3.2 \cdot 10^{12}$ V/m and $B_c= m_e\omega_0/e \approx 1.1 \cdot 10^4$ T. The particle density is normalized to the electron critical density $n_c \approx 1.1 \cdot 10^{21}$ cm$^{-3}$. $E_x$ represents the charge separation field and $E_y$ and $B_z$ are the fields of the electromagnetic waves. Because the laser pulse duration is long, the forward $E_{+}$ and backward  $E_{-}$ propagating components interfere in vacuum. We separate them according to the relations $E_{\pm}=\frac12 (E_y\pm B_z)$ which are exact in vacuum. We also use these relations in the plasma. Although they are not exact because the electromagnetic phase velocity in the plasma is different from $c$, we qualitatively check that they represent rather well the dynamics of both wave components.

In the representative case discussed in this article, a homogeneous hydrogen plasma slab with the initial density $n_{e0}=0.5\,n_c$ and the length  $l=150\,\lambda_0$ is located in the middle of a simulation box of $850\,\lambda_0$. The plasma is fully ionized and has a small initial temperature of 51 eV. The vacuum zones on the left and right sides allow free particle motion. The boundary conditions are absorbing for exiting fields and particles. The incident laser pulse has a linear p-polarization and a squared sinusoidal temporal shape with the maximum amplitude $a_0=2$ ($I_{\max}=5.5 \cdot 10^{18}$ W.cm$^{-2}$ ; $\lambda_0=1\,\mu$m). It enters in the simulation box at $t=0$ through the left boundary ($x=0$) and has a duration of $\tau_0 = 300\, T_0$. It reaches the plasma front at $t = 350\,T_0$. The simulation is stopped at $t_s \approx 1528\, T_0 \approx 5,1$ ps. This time is sufficient to follow all the plasma evolution linked to the production of energetic particles after the end of the laser pulse.

A quick estimate allows to show how much energy the particles may have. The laser energy equals $\int_{\tau_0} I_l(t)\dt \approx 2.7$ MJ/cm$^2$ and the total number of particles is $2n_{e0}l \approx 1.7 \cdot 10^{19}$ /cm$^2$. So, if all the laser energy is absorbed, the plasma would gain an average energy of 1.0 MeV per particle. Since we want to accelerate particles to high energies, we need to distribute the absorbed energy unequally, to activate the processes that allow to transfer a large amount of energy to a relatively small number of particles.

With these physical parameters, a very efficient laser energy absorption is reached by the plasma. It attains 68\% at the end of the simulation. About 38.5\% of the laser pulse energy is transferred to electrons and 29.5\% is communicated to ions. The reflected and transmitted laser energy are 27.7\% and 3.2\% of the incident energy, respectively. 

In this section, we discuss first the general characteristics of the plasma absorption process and then the spectral properties of excited electromagnetic and plasma waves.

\subsection{Plasma absorption}
\label{sec_plasmaabsorption}

\begin{figure}
\centering
\includegraphics[height=9cm]{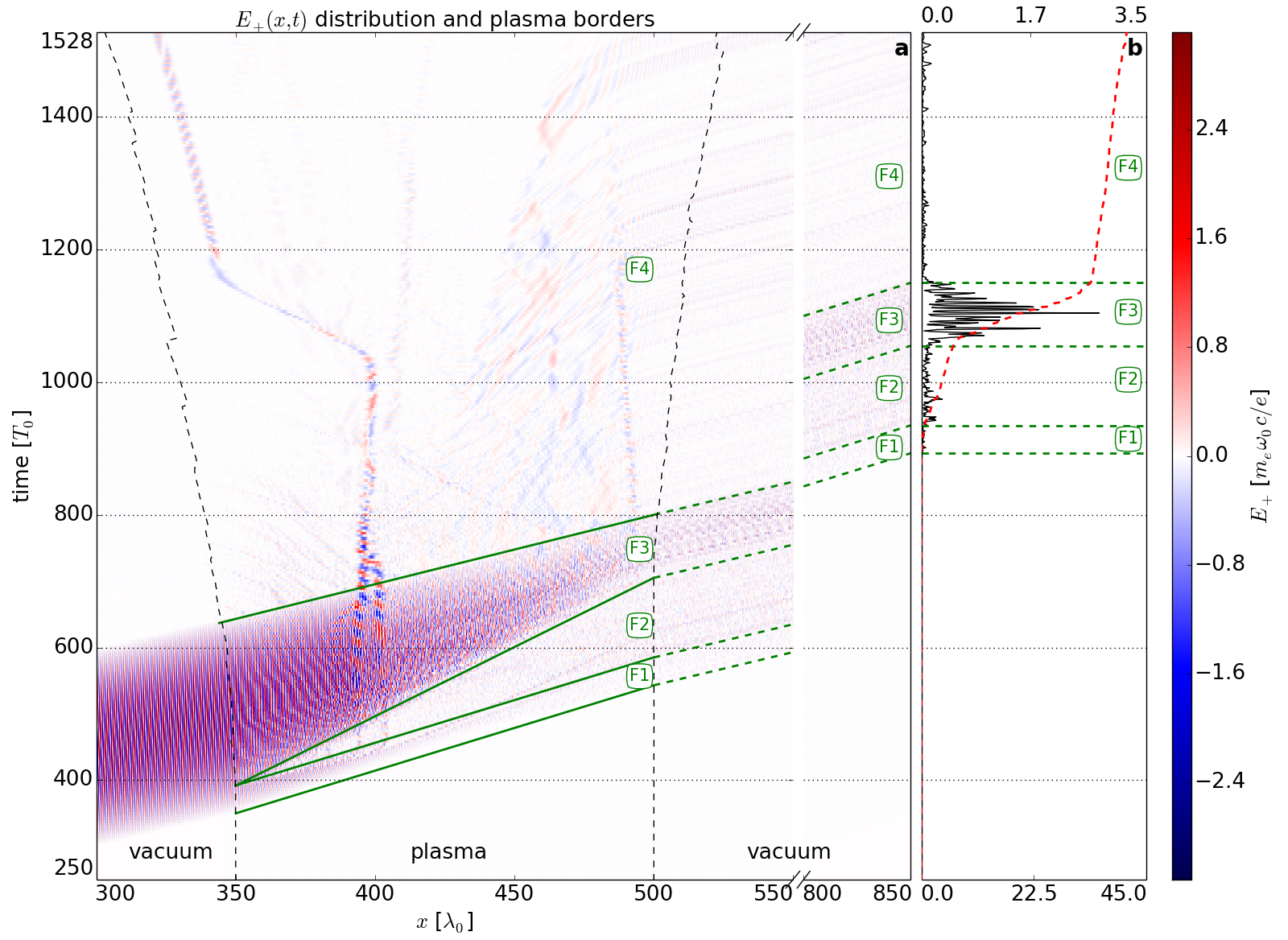} \\
\includegraphics[height=9cm]{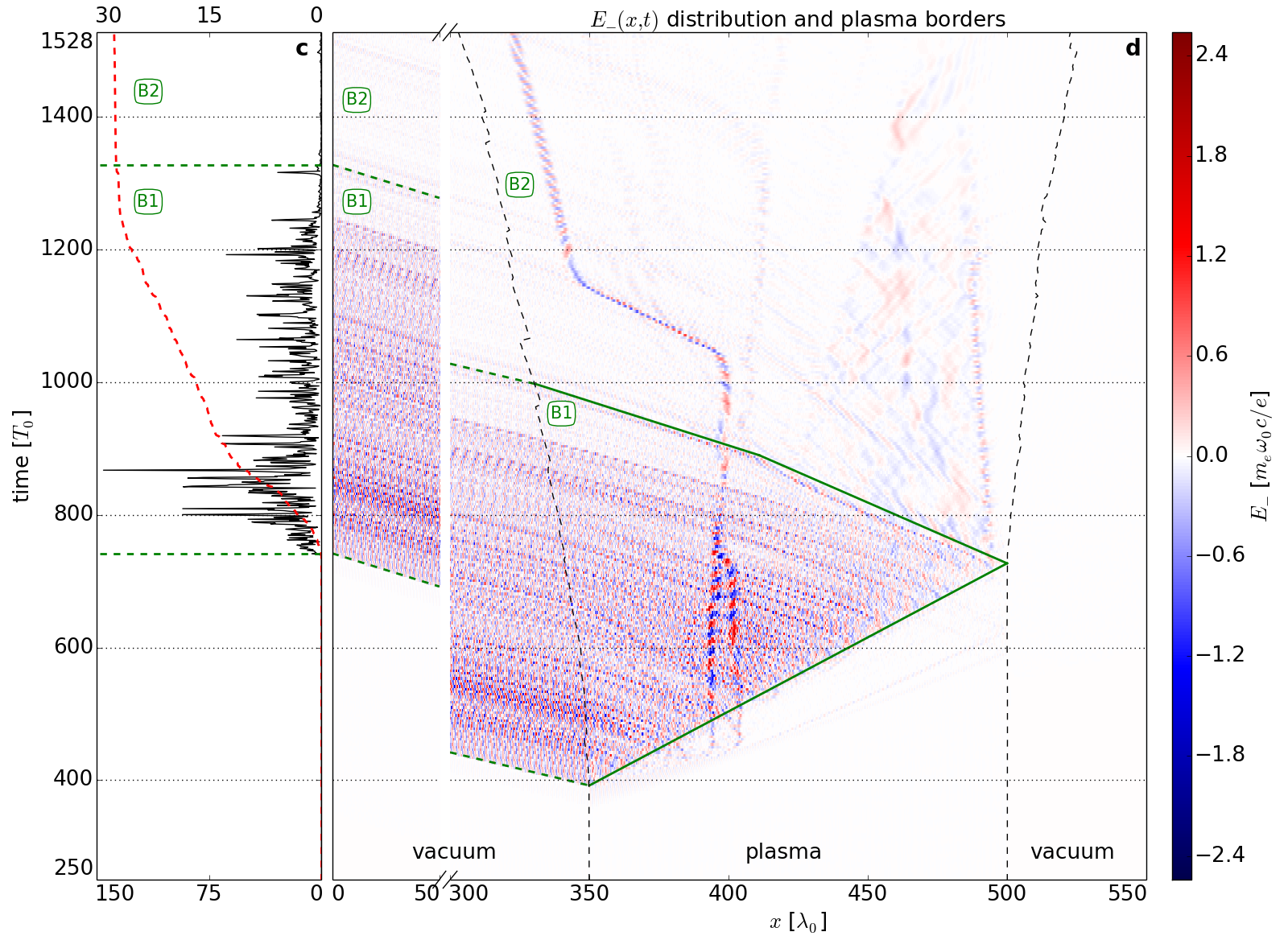}
\caption{Laser pulse interaction with a homogeneous plasma slab and Poynting fluxes measured at the left and right boundaries of the simulation box. Panels a and d give the amplitudes of forward $E_{+}$ and backward $E_{-}$ propagating waves, respectively, as a function of the longitudinal position $x$ and time $t$. Dashed lines delimit the plasma density $n_p$ where $n_p < 0.1\, n_c$. Panels b and c display the instantaneous (solid line) and cumulated (dashed red line) Poynting fluxes through the right (b) and left (c) boundaries of the simulation box.}
\label{fig_these-6_Eyright-trans}
\end{figure}

Figure \ref{fig_these-6_Eyright-trans} summarizes the temporal evolution of the electromagnetic waves in the plasma. Separation of the forward and backward propagating waves allows to identify the dominant nonlinear processes. The dashed black lines delimit the plasma boundaries, showing the plasma expansion during its interaction with the laser pulse. The forward and backward propagating waves are shown in Fig. \ref{fig_these-6_Eyright-trans}.a and Fig. \ref{fig_these-6_Eyright-trans}.d, respectively. The incident laser pulse propagates freely in vacuum and enters the plasma at $x = 350\, \lambda_0$, at the time $t=350\, T_0$. The tail of the laser pulse enters the plasma at $t \approx 650\, T_0$. During this interval, the plasma slightly expands. The interaction between the laser pulse and the plasma proceeds to the time $t \approx 800\, T_0$ when the tail of the pulse leaves the plasma.

Three main stages of laser plasma interaction are numbered in green in Fig. \ref{fig_these-6_Eyright-trans}.a. The first stage, corresponding approximately to the first $40\, T_0$ of the laser pulse, describes an approximately linear propagation of the front part of the pulse (Fig. \ref{fig_these-6_Eyright-trans}.a, zone (F1)). At the time $t \approx 390\,T_0$, when the laser pulse intensity is approximately one sixth of the maximum one ($I \sim 9 \cdot 10^{17}$ W.cm$^{-2}$), the parametric instability sets in and a strong backscattered wave is generated leading to almost complete extinction of the incident wave (see Fig. \ref{fig_these-6_Eyright-trans}.d).

The zone of the incident laser pulse extinction and the backscattered wave excitation extends inside the plasma with the velocity $\sim0.5\,c$ as it is represented by the upper boundary of zone (F2) in Fig. \ref{fig_these-6_Eyright-trans}.a and the lower boundary of zone (B1) in Fig. \ref{fig_these-6_Eyright-trans}.d. This extremely fast pump wave depletion process occurs over a length smaller than $10\,\lambda_0$ as discussed in Sec. \ref{sec_electronheating}. It is identified as the SRS process on the plasma waves modified by a relativistically intense incident laser pulse in Sec. \ref{sec_SRSconfirmation}.

There are large amplitude electron density oscillations left after the SRS coupling which are gradually transferring their energy to electrons. Zone (F3) in Fig. \ref{fig_these-6_Eyright-trans}.a shows the propagation of the remaining part of the laser pulse before it attains the SRS coupling zone. As it propagates through the zone of strong electrostatic plasma turbulence, the three-wave coupling is broken and the incident wave is attenuated much slower. This corresponds to the non-resonant interaction of a laser wave with strongly turbulent plasma waves. The tail of the laser pulse entering the plasma after the time $t \approx 600\, T_0$ is less absorbed and propagates almost with the vacuum light velocity. It is responsible for the major part of the transmitted light as it can be seen in Fig. \ref{fig_these-6_Eyright-trans}.b. 

In Fig. \ref{fig_these-6_Eyright-trans}.b, we present the Poynting flux computed at the right box boundary ($x=850\,\lambda_0$), either cumulated over time steps of a $1\, T_0$ (solid curve) or cumulated over all the simulation duration (dashed curve). The $1T_0$-cumulated transmitted flux is normalized to the energy of $1\, T_0$ laser step where $I_l=I_{\max}/2$. The cumulated transmitted flux is normalized to the total laser pulse energy.

We distinguish three regimes of transmission in Fig. \ref{fig_these-6_Eyright-trans}.b. At the time intervals (F1) and (F2), the transmission is very low, it represents $\sim 0.5$\% of the total laser pulse energy. The interval (F3) corresponds to an enhanced transmission of the laser pulse tail from $t \approx 890$ to $1060\,T_0$. This accounts for 2.2\% of the total laser pulse energy. As it is shown in Sec. \ref{sec_spectralaanalysis}, the frequency of the transmitted light is approximately equal to the laser frequency $\omega_0$. After the laser pulse ends, there is still emission of small amplitude electromagnetic waves, which corresponds to the plasma radiation (see zone (F4) in Fig. \ref{fig_these-6_Eyright-trans}.b). Finally, the accumulated plasma transmission reaches 3.2\% at the end of the simulation.

Two regimes of backscattered radiation can be seen in Fig. \ref{fig_these-6_Eyright-trans}.d. The waves are created in the SRS coupling zone just at the lower boundary of zone (B1). They are propagating almost freely through the zone of strong plasma wave turbulence and reaching the left box boundary at $t\approx 740\,T_0$. The total duration of the reflected pulse is about $600\,T_0$, almost twice the duration of the incident pulse. With exception of an intense transient spike with a duration of $130\, T_0$ at the beginning of the reflected pulse, its intensity is approximately constant corresponding to an average instantaneous reflectivity of the order of $\sim 20$\% (see Fig. \ref{fig_these-6_Eyright-trans}.c). The reflected pulse contains $\sim27$\% of the incident laser energy. One third of it is emitted during the transient stage and two thirds during the permanent stage. There is also emission of weak electromagnetic waves after the end of the main reflected pulse at $t \approx 1330\,T_0$. The intensity of these waves accounts for less than 0.4\% of the laser pulse energy. These waves are of the same origin as  the post emission in the forward direction for $t > 1150\,T_0$. This is further confirmed by the spectral analysis in Sec. \ref{sec_spectralaanalysis}.

In conclusion, the absorption of the laser pulse by the plasma proceeds in the following steps. The front of the laser pulse penetrates in the undisturbed plasma. The laser pulse is progressively attenuated due to the excitation of the backscattered wave. The SRS zone extends inside the plasma with the velocity $\sim 0.5\,c$ leaving behind it a strongly turbulent plasma where the laser absorption is strongly reduced. Therefore only the trailing part of the laser pulse succeeds to travel across the plasma. After the laser pulse leaves the plasma, there are still electromagnetic waves trapped in it (see zone 4 in Fig. \ref{fig_these-6_Eyright-trans}.a). These waves have however much smaller amplitudes and decay on the time scale of $300-400\,T_0$ as the plasma expands.

There are also particular long living objects seen at the position $x\sim 400\,\lambda_0$ almost from the beginning of the interaction process. These are electromagnetic plasma cavities \cite{kim1974}.

\subsection{Spectral analysis of the electromagnetic and plasma waves}
\label{sec_spectralaanalysis}

\begin{figure}
\centering
\includegraphics[width=16cm]{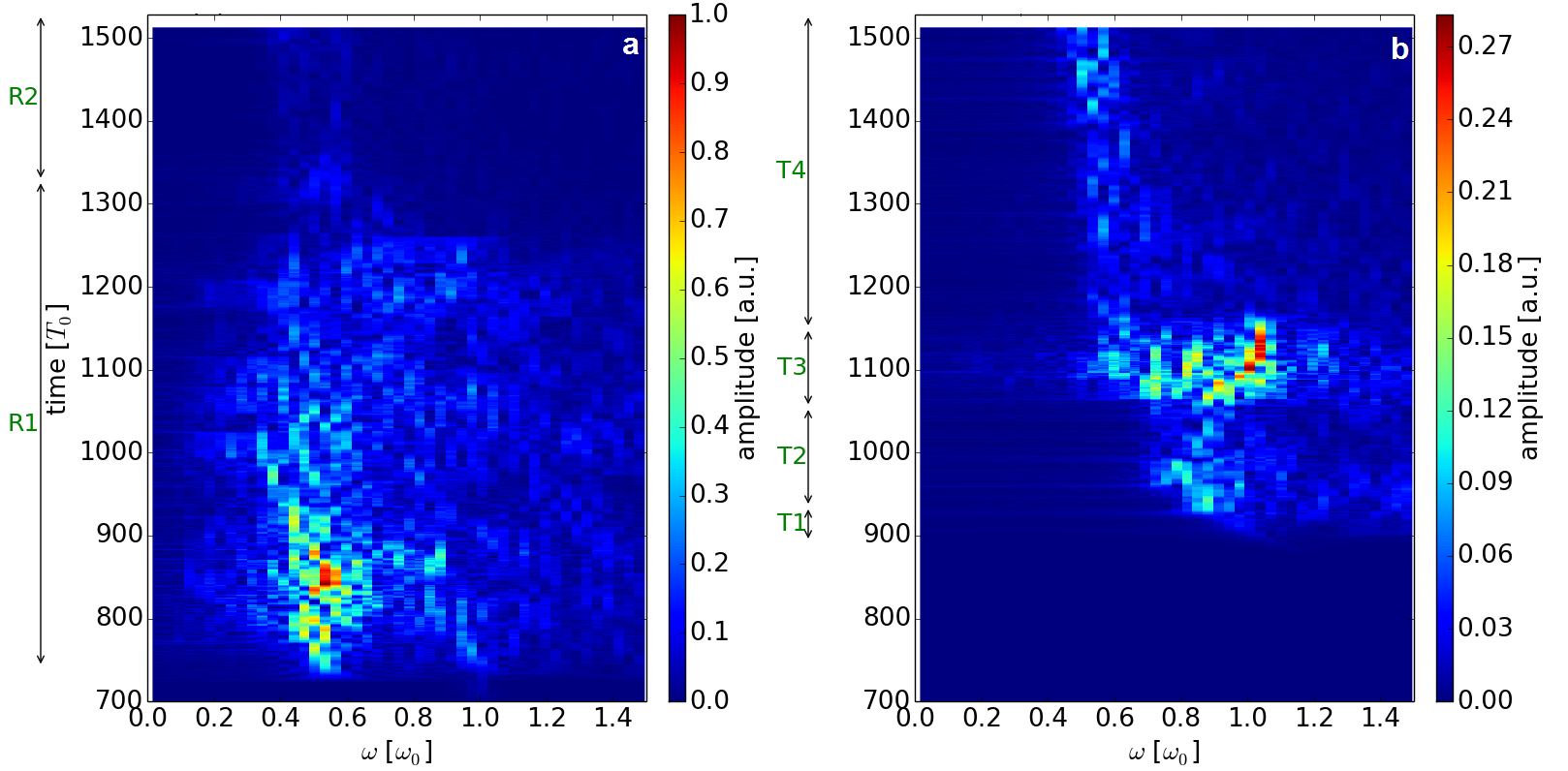}
\caption{Time-frequency analysis of the electromagnetic field leaving out the left boundary (a) and the right boundary (b) of the simulation box. Both panels are normalized to the same scale.}
\label{fig_these-6_timefreqanalysis_0-850}
\end{figure}

The SRS origin of the backscattered emission is confirmed by the spectral analysis of the waves in the plasma. Figure \ref{fig_these-6_timefreqanalysis_0-850} presents the time frequency analysis of the backscattered and transmitted fields measured in vacuum at the left and right boundaries of the simulation box, respectively.

The spectral analysis was performed with a Fast Fourier Transform (FFT) of electric fields at a fixed spatial position. The FFT was computed over a time window $\tau_\omega = N\Delta t \approx 31.8\, T_0$ ($N=4000$) moving along the time axis. This method allows us to observe the time evolution of wave frequencies excited in the plasma. The accuracy of the temporal evolution of the spectra is limited by the time window width $\frac12\tau_\omega$ of the FFT. The frequency resolution is $\Delta \omega = 2\pi/N \Delta t \approx 0.03\,\omega_0$. 

The dominant  frequency of reflected electromagnetic waves in Fig. \ref{fig_these-6_timefreqanalysis_0-850}.a is equal to $\sim 0.53\,\omega_0$. It goes through the box left boundary from $t=740\, T_0$ to $924\, T_0$. Its amplitude increases with time until $t \approx 880\, T_0$. Simultaneously the signal bandwidth increases and reaches $\sim 0.28\,\omega_0$. Then, the field amplitude decreases until $t \approx 1330\, T_0$. This time interval from 740 to  $1320\, T_0$ corresponds to the scattering zone (B1) in Fig. \ref{fig_these-6_Eyright-trans}.d. The weak signal continues for later times at approximately the same frequency of $0.5\,\omega_0$. It corresponds to zone (B2) in Figs. \ref{fig_these-6_Eyright-trans}.c-d. It is shown in Sec. \ref{sec_SRSconfirmation} that these backscattered electromagnetic waves originate from the SRS parametric instability.

Moreover, Fig. \ref{fig_these-6_timefreqanalysis_0-850}.a shows that only a small part of the back-scattered waves have a frequency equal to the laser frequency $\omega_0$. This indicates that, for the interaction parameters used in this simulation, there is almost no laser reflection. Then, the ponderomotive force at the front side of the plasma is not sufficiently strong to produce the electron pileup, which leads to laser reflection and so harmful losses of energy. 

Figure \ref{fig_these-6_timefreqanalysis_0-850}.b shows the time frequency analysis of the forward propagating field $E_{+}$ measured at the right boundary of the simulation box. It can be compared with Fig. \ref{fig_these-6_Eyright-trans}.a-b. The signal reaches the right border at $t \approx 940\, T_0$ and is stopped around $\sim 1150\, T_0$ which corresponds to the tail of the laser pulse. This clearly shows a shortening of the pulse duration due to the plasma absorption. Its frequency broadens from 0.8 to $1.05\,\omega_0$ due to scattering on turbulent plasma density fluctuations. At later times $t> 1150\,T_0$ the plasma emits weak low frequency waves in the interval $(0.5-0.6)\, \omega_0$.

\begin{figure}
\centering
\includegraphics[scale=0.35]{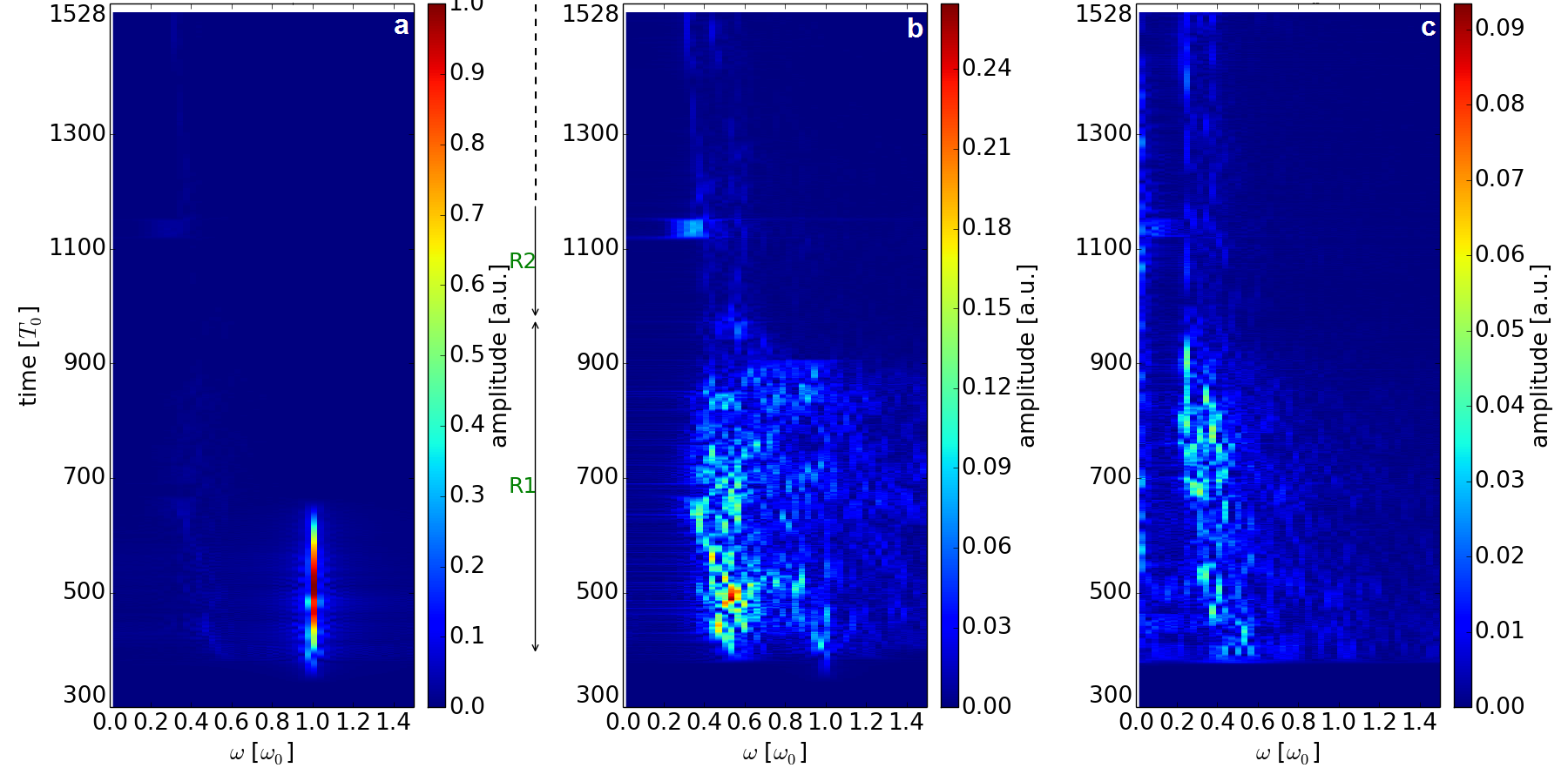}
\caption{Time-frequency analysis of the forward-propagating $E_+$ (a), backward-propagating $E_{-}$ (b) and electrostatic $E_x$ (c) fields at the distance of $3\, \lambda_0$ from the left plasma edge ($x=353\, \lambda_0$). All three panels are normalized using the same scale.}
\label{fig_these-6_timefreqanalysis_353}
\end{figure}

The origin of the frequency shift of the transmitted and backscattered waves can be understood by considering the spectra of electromagnetic and electrostatic waves inside the plasma. Figure \ref{fig_these-6_timefreqanalysis_353} displays the time-frequency analysis of the forward- and backward-propagating electromagnetic waves and electrostatic $E_x$ fields measured at the plasma front ($x=353\, \lambda_0$). One should notice that the time-frequency analysis of the backscattered wave inside the plasma (Fig. \ref{fig_these-6_timefreqanalysis_353}.b) and inside vacuum (Fig.  \ref{fig_these-6_timefreqanalysis_0-850}.a) are very similar. This confirms the viability of the separation of the forward and backward fields $E_\pm$ in the plasma via the relation $E_\pm = \frac12 (E_y\pm B_z)$.

The forward propagating wave in the plasma (Fig. \ref{fig_these-6_timefreqanalysis_353}.a) has a narrow spectrum centred at the pump frequency $\omega \approx \omega_0$. Its duration corresponds to the incident pulse interacting with the front part of the plasma. According to Fig. \ref{fig_these-6_Eyright-trans}.a, the laser pulse is not yet depleted at this point.

The backscattered electromagnetic wave, shown in Fig. \ref{fig_these-6_timefreqanalysis_353}.b, has a frequency equal to $0.53\, \omega_0$. It appears with a short delay less than $40\,T_0$ with respect to the pump arrival and has a much longer duration of about $600\, T_0$. This is in agreement with the duration of the backscattered signal observed in Figs. \ref{fig_these-6_Eyright-trans}.c-d. In both analysis, we observe that a back-scattered wave is emitted by the plasma approximately $40\, T_0$ after the beginning of the interaction (zone (B1)). The amplitude and the bandwidth of backscattered waves increase with time over the first 100 periods and then decrease when the laser pulse goes out the plasma. 

Fig. \ref{fig_these-6_timefreqanalysis_353}.c shows the development of an electrostatic wave. Its time evolution and  duration match well the backscattered wave. Initially, the electrostatic wave frequency is equal to $0.53\, \omega_0$. Then it decreases to $0.24\, \omega_0$. The sum of the plasma and backscattered wave frequencies at the time interval 400-450 $T_0$ matches well the laser frequency thus indicating the resonant three-wave process.

\section{SRS in a near-critical plasma}
\label{sec_SRSconfirmation}

The frequency matching:
\begin{equation}\label{eq3}
\omega_0 = \omega_s + \omega_p,
\end{equation}
where $\omega_0$ , $\omega_s$ and $\omega_p$ are the laser frequency, the scattered wave frequency and the plasma frequency respectively, may correspond to the stimulated Raman scattering (SRS). In this section, we show that these waves also verify the wave vector matching and explain the reason why such an instability can be excited in a plasma with a density significantly higher than the quarter critical density, which usually does not permit the propagation of the scattered wave.

\begin{figure}
\centering
\includegraphics[scale=0.35]{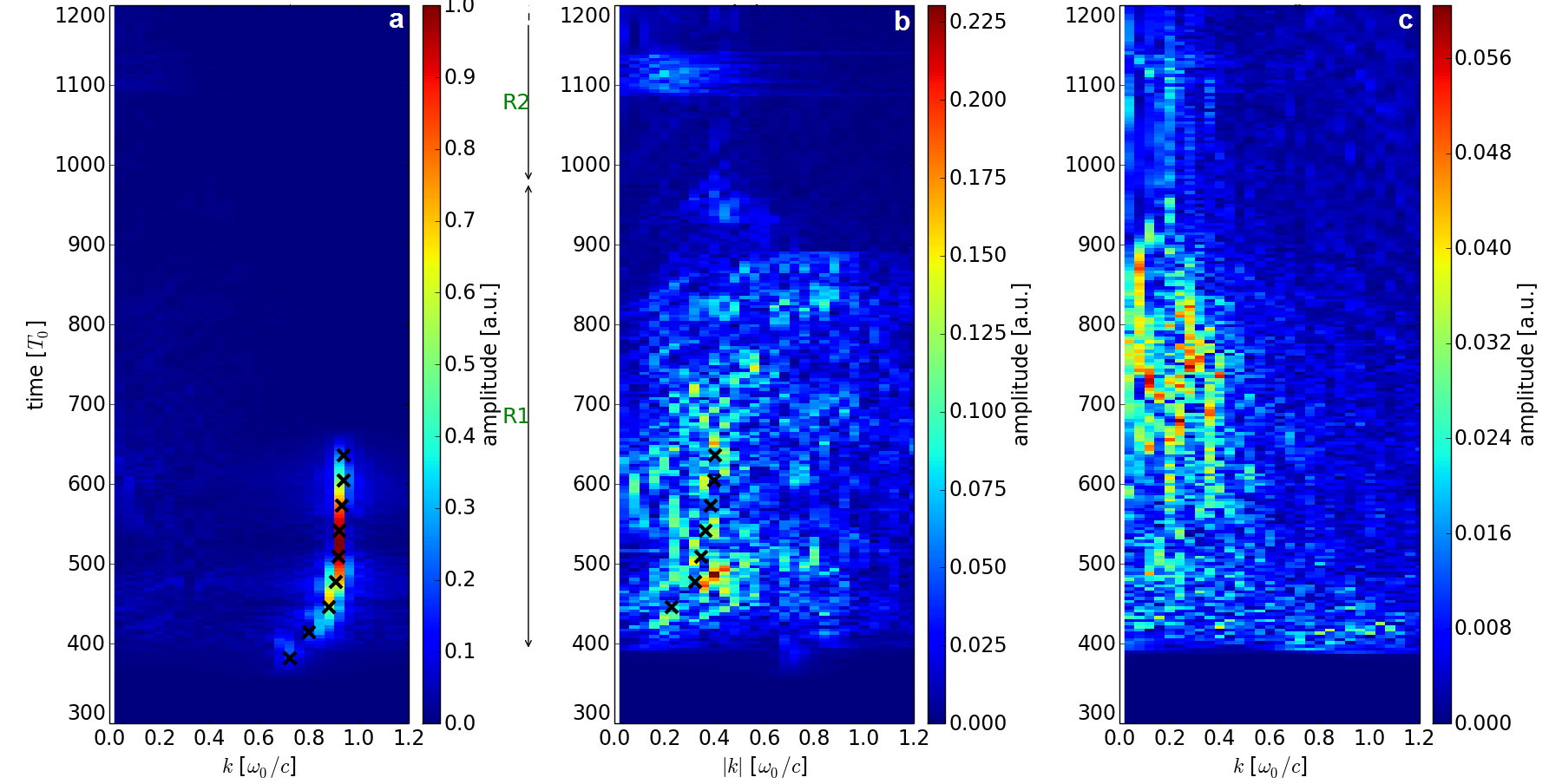}
\caption{Time evolution of the spatial spectra of the forward-propagating $E_+$ (a), backward-propagating $E_{-}$ (b) and electrostatic $E_x$ (c) fields in the plasma, in the interval (350-375) $\lambda_0$. All three panels are normalized using the same scale. The black crosses in panels (a) and (b) give the theoretical values of the wave number $k_0$ and $|k_s|$ in the plasma of the laser wave ($\omega_0$) and the scattered wave ($\omega_s=0.53\omega_0$) respectively.}
\label{fig_these-6_timekanalysis_350-375}
\end{figure}

Figure \ref{fig_these-6_timekanalysis_350-375} displays the temporal evolution of the spatial spectra for the forward- and backward-propagating electromagnetic and electrostatic $E_x$ fields measured at the plasma front. It is computed along a fixed interval (from $x=350\, \lambda_0$ to $x=375\, \lambda_0$). 

The forward propagating wave (Fig. \ref{fig_these-6_timekanalysis_350-375}.a) has a narrow spectrum centred at the wave number $k_0$ of the laser pulse in the plasma. We measure that it increases from $0.72\,\omega_0/c$ to $0.92\,\omega_0/c$ until $t\approx 650\, T_0$. It can be demonstrated that this wave number $k_0$ verifies the relativistic dispersion relation of an electromagnetic wave in the plasma:
\begin{equation}\label{eq4} 
\omega_0^2 = \frac{\omega_p^2}{\langle\gamma_0\rangle} + k_0^2c^2
\end{equation}
where $\langle\gamma_0\rangle$ is the average relativistic factor of an electron in the field of a linearly polarized wave \cite{gibbon2005}. This relation shows that the wave number of the electromagnetic wave increases with the electron energy $\langle\gamma_0\rangle$. Indeed, we measure, in our simulation, that the electron mean relativistic factor $\langle\gamma_0\rangle^{PIC}$, computed in the same interval $x=(350$-$375)\, \lambda_0$, also increases with time. This is further confirmed by the comparison of the spatial spectra of the forward-propagating $E_+$ with the theoretical values (black crosses) of $k_0^{th} = \frac{1}{c}\sqrt{\omega_0^2-\omega_p^2/\langle\gamma_0\rangle^{PIC}}$, where $\omega_p^2=\frac{n_e}{n_c}\omega_0^2$ and $\langle\gamma_0\rangle^{PIC}$ is computed in the plasma front ($x=(350$-$375)\, \lambda_0$) periodically. This comparison gives a very good agreement.

The time evolution of the absolute value of the wave number $|k_s|$ of the backscattered electromagnetic waves is shown in Fig. \ref{fig_these-6_timekanalysis_350-375}.b. It is quite large and, in agreement with Figs. \ref{fig_these-6_Eyright-trans} and \ref{fig_these-6_timefreqanalysis_353}, it has a duration two times longer than the pump wave. Before reaching the left boundary of the simulation box, these waves travel through the edge of the plasma front where the absolute value of their wave number $|k_s|$, according to Fig \ref{fig_these-6_timekanalysis_350-375}.b, increases from $0.2\, \omega_0/c$ to $0.4\, \omega_0/c$. As for the pump wave, this shift is due to the rise of the electron energy in the plasma since the scattered waves also verify the dispersion relation \eqref{eq4}. We show that it is in good agreement with the theoretical values (black crosses) of $|k_s|^{th} = \frac{1}{c}\sqrt{\omega_s^2-\frac{\omega_p^2}{\langle\gamma_0\rangle}}$, with $\omega_p^2=\frac{n_e}{n_c}\omega_0^2$ and $\omega_s=0.53\,\omega_0$. Then, the dominant scattered wave, whose frequency equals $\omega_s=0.53\,\omega_0$, observed in vacuum and in the plasma (see Sec. \ref{sec_spectralaanalysis}) also appears in the spatial spectra.

Similarly to the time-frequency analysis (Fig \ref{fig_these-6_timefreqanalysis_353}.c), the evolution of the spatial spectrum of the electrostatic wave in the plasma is shown in Fig. \ref{fig_these-6_timekanalysis_350-375}.c. Its wave number initially (at $t \approx 400\, T_0$) equals $\sim 1.0\, \omega_0/c$. This initial wave quickly disappears producing then a large spectrum with low wave numbers corresponding to a broken highly non linear electrostatic wave.

Hence, initially, at $t \approx 400\, T_0$, when the instability develops at the front edge of the plasma, the combination of the back-scattered $k_s \approx -0.2\, \omega_0/c$ and plasma wave $k_p \approx 1.0\, \omega_0/c$ wave numbers matches the wave number of the incident wave in the plasma $k_0 \approx 0.8\, \omega_0/c$:
\begin{equation}\label{eq5} 
k_0 = k_s + k_p.
\end{equation}
Thus, the spatial and temporal spectral analysis confirms that there are three waves coupled in the plasma. This corresponds to the Stimulated Raman Scattering instability as the incident and backscattered waves verify the relativistic dispersion relation \eqref{eq4} for electromagnetic waves and the third wave corresponds to the dispersion relation of the plasma wave $\omega_p \approx \omega_{p0}/\sqrt{\langle\gamma_0\rangle}$. Fig. \ref{fig_these-6_timekanalysis_350-375} shows that this three wave coupling only exists for a few tens $T_0$. We have repeated this temporal and spatial analysis at several positions of the plasma confirming that this instability develops all along the plasma while the laser wave has a sufficiently high amplitude. 

It has been already demonstrated analytically that the SRS instability could be excited  in a plasma with a density significantly higher that the quarter critical density for relativistic laser intensity \cite{guerin1995}. This is explained by the relativistic increase of the effective mass of electrons oscillating in a large amplitude laser wave and the corresponding decrease of the effective plasma frequency $\omega_p/\sqrt{\gamma_e}$. This allows for a scattered electromagnetic wave to be produced by the SRS process and propagate in plasma thanks to the relativistic self-induced transparency.

The SRS dispersion relation was obtained by S. Gu\'erin et al. \cite{guerin1995} considering the instability of a circularly polarized wave in a cold plasma:
\begin{equation}\label{eq6} 
D_+D_-=\frac{\omega_{p0}^2a_0^2}{4\gamma_0^3}\left(\frac{k_p^2c^2}{D_p}-1\right)\left(D_++D_-\right) 
\end{equation}
where $D_p$ and $D_\pm$ correspond, respectively, to the dispersion relation of the electron plasma wave and the electromagnetic waves:
\begin{align}\label{eq7} 
D_p &= \omega_p^2-\frac{\omega_{p0}^2}{\gamma_0} & D_\pm &= \omega_p^2 - k_p^2c^2 \pm 2(\omega_0\omega_p-k_0k_pc^2)
\end{align}
where $\omega_{p0}^2/\omega_0^2 = n_{e0}/n_c$ and $\gamma_0 = \sqrt{1+a_0^2/2}$.

We apply this dispersion relation for a linearly polarized wave by replacing $\gamma_0$ by $\langle \gamma_0 \rangle$ and solve it for the parameters of the simulation presented above ($n_e = 0.5\, n_c$ ; $a_0=2$). The solutions $\omega_p = \Re \omega_p + \i \Im \omega_p$ verifying $\Gamma = \Im \omega_p >0$ correspond to the unstable electrostatic modes that may correspond to the SRS instability. We indeed found that, for these interaction parameters, unstable solutions $\Im \omega_p>0$ exist so that the SRS instability can appear in the plasma for this laser pulse intensity. The maximum growth rate $\Gamma$ of the instability is $\approx 0.35\, \omega_0$ when the laser pulse reaches its maximum intensity ($a_0=2$).

\begin{figure}
\centering
\includegraphics[scale=0.35]{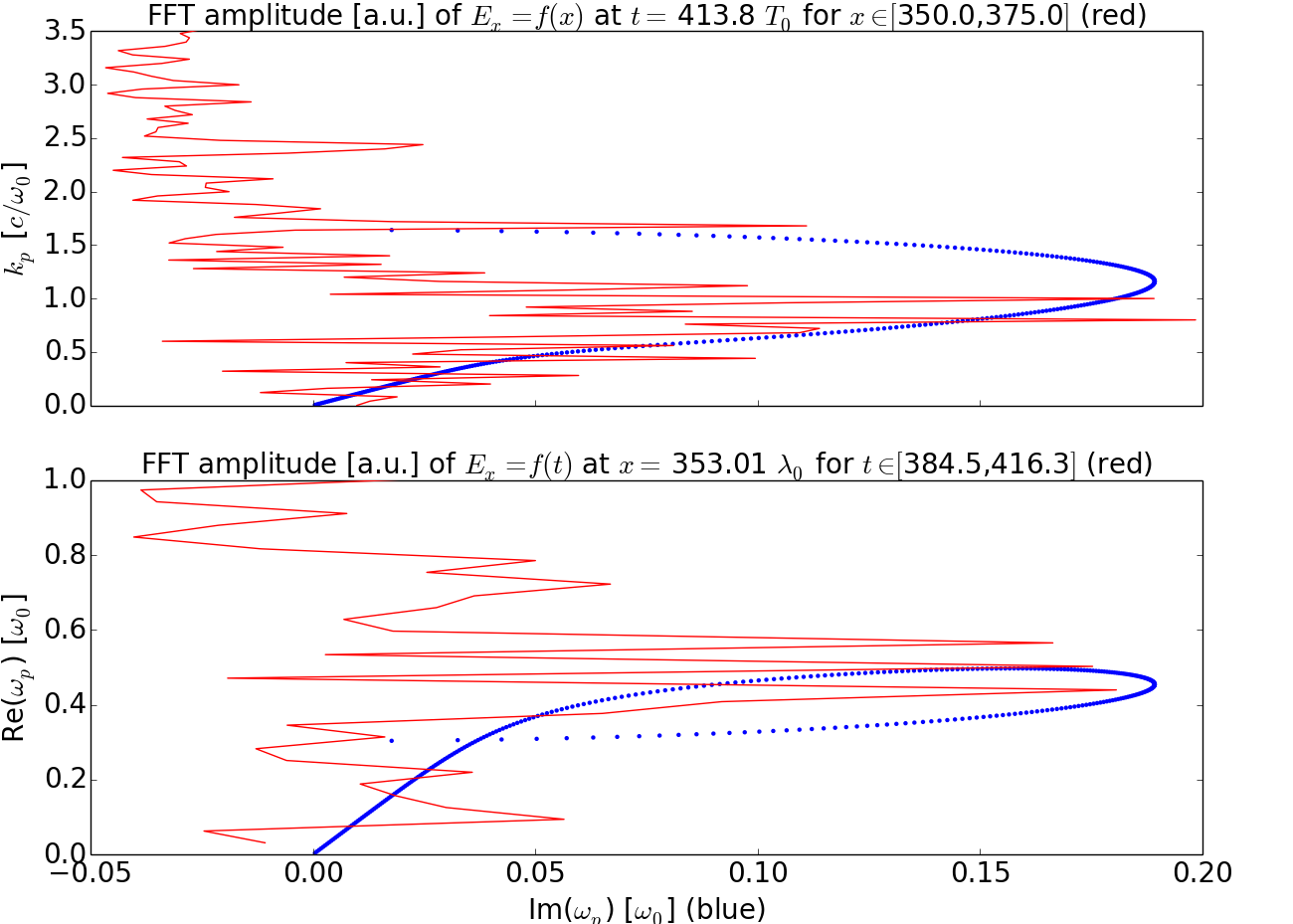}
\caption{Comparison of the electrostatic wave number $k_p$ (a) and frequency $\Re\omega_p$ (b) as a function of the growth rate $\Gamma=\Im \omega_p$ from the solution of Eq. \eqref{eq6} (blue curve) and as a function of FFT amplitude from simulation analysis (red curve) in the plasma range $x=(350$-$375)\, \lambda_0$ and at $t \approx 414\, T_0$.}
\label{fig_these-6_compar_guerin}
\end{figure}

In order to compare the predictions of the theoretical model with simulation results, we present, in Fig. \ref{fig_these-6_compar_guerin}, the electrostatic wave frequency $\Re\omega_p$ and the wave number $k_p$ as a function of the growth rate $\Gamma$ (blue) and as a function of FFT amplitude from simulation analysis (red). The temporal and spatial Fourier analysis of the electrostatic fields computed at the plasma front, in the simulation, shows that the electrostatic wave with the frequency $\omega_p = (0.50 \pm 0.08) \,\omega_0$ and the wave number $k_p = (1.0 \pm 0.1) \,\omega_0/c$ is the dominant mode in the plasma. We compare these results with the solution of Eq. \eqref{eq6}  calculated for the laser amplitude $a_0=1$ and the electron energy $\langle \gamma_0 \rangle = 1.38$ measured at the plasma front, where the FFT of the electrostatic field was computed (see Fig. \ref{fig_these-6_compar_guerin}). This comparison shows that the dominant mode of the plasma corresponds to $\Gamma \sim 0.15\,\omega_0$ which is close to the maximum growth rate $\Gamma$ predicted by Eq. \eqref{eq6}. Thus, the main electrostatic wave observed in the plasma corresponds to the mode having one of the highest probability to be excited in the plasma.

Thus, a comparison with an analytical model confirms excitation of the SRS instability, and excitation of fast growing electrostatic waves. The wave number analysis shows, at each spatial position, that this three waves coupling exists for a short time of a few tens of laser periods. It is quickly broken, the plasma wave amplitude saturated and the spectrum extends to small wave numbers. Then the incident laser wave may propagate deeper in plasma and excite the SRS in a fresh plasma layer.

\section{Electron heating and ion acceleration}
\label{sec_electronheating}

The laser energy deposited in the plasma waves is further transferred to electrons after the spectrum broadening and wave-breaking of the plasma waves. In this section, we study the electron heating and the ion acceleration.

\begin{figure}
\centering
\includegraphics[scale=0.4]{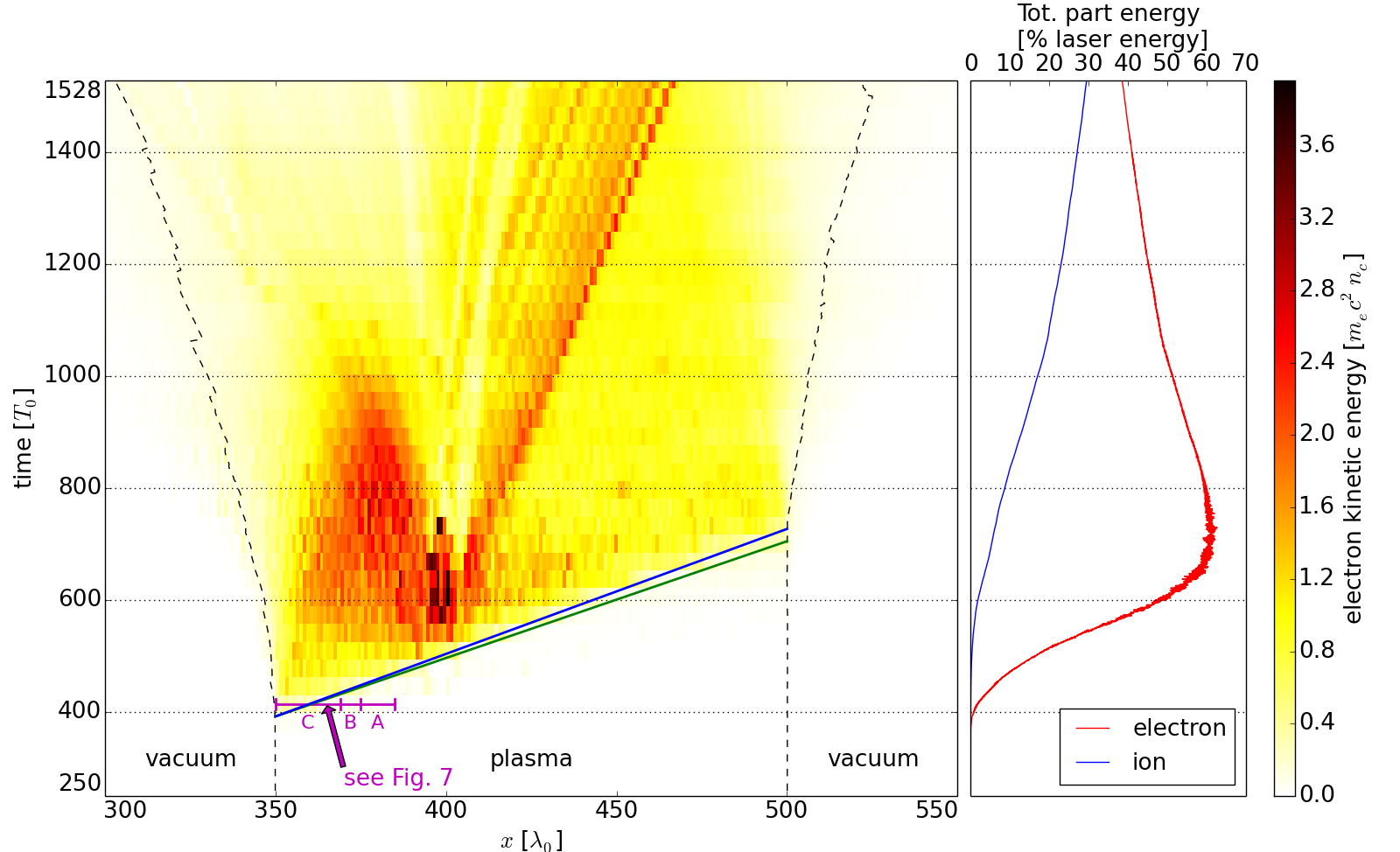}
\caption{Electron energy density as a function of space and time (a) and cumulated electron (red) and ion (blue) energy (b) as a function of time. Dashed lines delimit the plasma density $n_p$ where $n_p < 0.1\, n_c$. The green line (corresponding to the (2)-(3) boundary line in Fig. 1.a) shows the arrival of the high amplitude laser pulse. The blue line (corresponding to the (0)-(1) boundary line in Fig. 1.d) shows the triggering of the back-scattered electromagnetic waves by the plasma.}
\label{fig_these-6_electronheating}
\end{figure}

Figure \ref{fig_these-6_electronheating}.a displays the electron energy density as a function of time and space. It is computed by calculating the total electron energy over segments of 1 $\lambda_0$, each $31.8\, T_0$. The electrons absorb energy all along the plasma over the time interval from $t=400\, T_0$ to $t=700\, T_0$. At each plasma position, we observe that electrons are heated once the SRS instability is triggered. This is confirmed by Fig. \ref{fig_these-6_electronheating}.b, which displays the total electron energy of the plasma as a function of time. It shows that the electron energy increases approximately linearly from $t=400\, T_0$ to $t=700\, T_0$ where it reaches approximately 61 \% of the total laser pulse energy.

The correlation of the time of the electron energy gain and the time of SRS instability development can be confirmed further by comparison of Fig. \ref{fig_these-6_electronheating} to Fig. \ref{fig_these-6_Eyright-trans}. The arrival of the high amplitude part of the laser pulse and the onset of electromagnetic waves back-scattering are presented in Fig. \ref{fig_these-6_electronheating}.a, by the green line (corresponding to the (2)-(3) boundary line in Fig. \ref{fig_these-6_Eyright-trans}.a) and the blue line (corresponding to the (0)-(1) boundary line in Fig. \ref{fig_these-6_Eyright-trans}.d), respectively. Figure \ref{fig_these-6_electronheating}.a shows that the electron heating occurs just after the development of the SRS instability. This confirms that the SRS instability and the subsequent plasma wave breaking are indeed the origin of the strong electron heating. This process occurs over a duration of less than $100\, T_0$.

\begin{figure}
\centering
\includegraphics[scale=0.4]{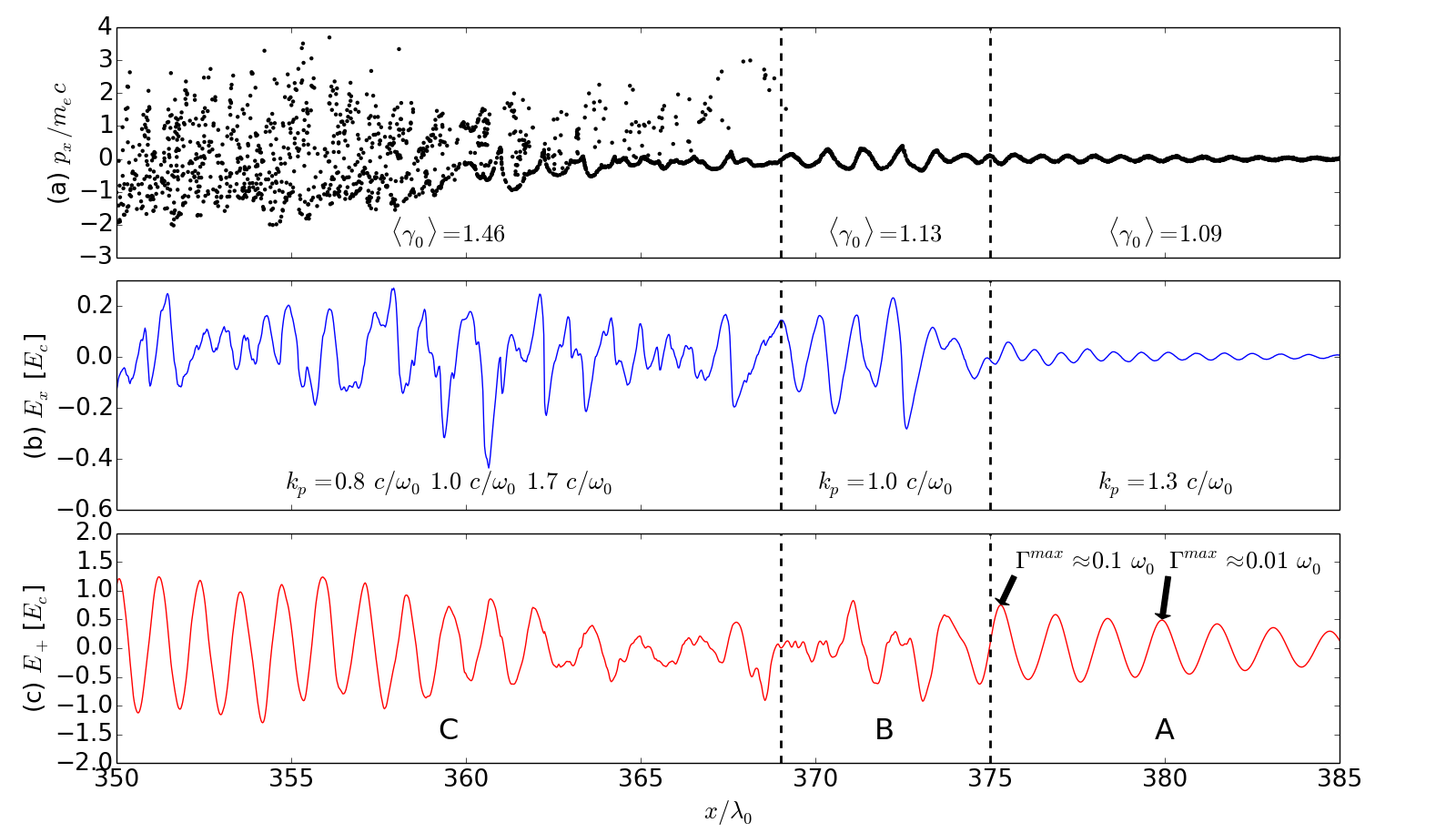}
\caption{Longitudinal electron phase-space (a), electrostatic field $E_x$ (b) and forward propagating field $E_{y,for}$ at the plasma front ($350$-$385\,\lambda_0$) at $t \approx 414\, T_0$.}
\label{fig_these-6_phasespace_electron}
\end{figure}

This description of the interaction of the laser pulse with the plasma and the electron heating is confirmed by analysing the phase-space of the electrons and the field distribution along the interval shown in magenta in Fig. \ref{fig_these-6_electronheating} at $t \approx 414\, T_0$. In Figure \ref{fig_these-6_phasespace_electron}, we present the longitudinal electron phase-space (a), the electrostatic field (b) and the forward propagating field $E_+$ as a function of space in the interval $x=(350$-$385)\,\lambda_0$ at $t \approx 414\, T_0$. We distinguish three zones of wave-particle interaction (see areas A, B and C). As the laser pulse propagates in the direction of increasing $x$ with a rising amplitude, the interaction time goes from the right to the left in this figure.

In zone A, the electrons oscillate in the field of the front part of the laser pulse with an increasing energy as the laser pulse amplitude increases (see Figs. \ref{fig_these-6_phasespace_electron}.a and \ref{fig_these-6_phasespace_electron}.c). The mean electron relativistic factor $\langle\gamma_0\rangle$ equals 1.09 in this zone. These oscillations form a low amplitude plasma wave whose wave number equals $k_p \approx 1.3\, \omega_0/c$ (Figs. \ref{fig_these-6_phasespace_electron}.b). By solving the dispersion relation \eqref{eq6} for the laser amplitude and the electron energy measured at $x=380\,\lambda_0$, we calculate that the growth rate of the SRS instability at that moment attains the value of $0.01\,\omega_0$ so the instability has not yet set in.

However, at $x=375\,\lambda_0$, the maximum growth rate $\Gamma^{max}\approx 0.1\, \omega_0$ is ten times higher than in $x=380\,\lambda_0$ so that the SRS instability is excited. We thus observe in zone B, the apparition of a high amplitude electrostatic wave with the number equal to $k_p \approx 1\, \omega_0/c$ (Fig. \ref{fig_these-6_phasespace_electron}.b). Indeed, we have shown in Sec. \ref{sec_SRSconfirmation} that this wave number corresponds to the mode excited by the SRS instability. Figure \ref{fig_these-6_phasespace_electron}.a shows that electrons oscillate in phase with the excited plasma wave. Nevertheless, their mean relativistic factor remains relatively small, $\langle\gamma_0\rangle = 1.13$.

In zone C, the electrons have escaped from the plasma wave. Their longitudinal momentum $p_x/me_c$ has largely increased (Fig. \ref{fig_these-6_phasespace_electron}.a) and their mean kinetic energy $(\langle\gamma_0\rangle -1)m_ec^2$ reaches a level of $0.46\,m_ec^2$ (0.23 MeV). The electrostatic wave is strongly non-linear and contains several strong modes $k_p=0.8\, \omega_0/c$, $1.0\, \omega_0/c$ and $1.7\, \omega_0/c$. According to Fig. \ref{fig_these-6_timekanalysis_350-375}, the mode $k_p \approx 1\, \omega_0/c$ excited by the SRS instability only exists for a short time. It is then replaced by the low modes of highly non linear plasma waves. Figure \ref{fig_these-6_phasespace_electron}.c shows damping of the laser wave amplitude in the zone B and in the beginning of zone C. This observation confirms that the laser pulse absorption occurs over a short length of approximately $10\, \lambda_0$, as discussed in Sec. \ref{sec_plasmaabsorption}.

Finally, from $x=350\;\lambda_0$ to $x=363\;\lambda_0$, we see in Fig. \ref{fig_these-6_phasespace_electron}.a that the plasma has became very turbulent so the laser pulse propagates in the plasma without being absorbed but just being modulated (see Fig. \ref{fig_these-6_phasespace_electron}.c). This free propagation of the laser pulse is also observed in zone (F3) in Fig. \ref{fig_these-6_Eyright-trans}.a.

Besides, Fig. \ref{fig_these-6_electronheating} shows that after the blue-green lines region where the SRS instability takes place, there is also a significant electron heating near the cavity position ($x\sim400\, \lambda_0$). The hot electrons are partially trapped in the cavity and slowly spread over the plasma during the time interval between $t \sim 500\, T_0$ to $t \sim 1100\, T_0$. The energy density of the electrons trapped in the cavities is 2-3 times larger than the average electron energy. However, their relative number is small and does not affect the overall energy balance in the plasma.

\begin{figure}
\centering
\includegraphics[scale=0.32]{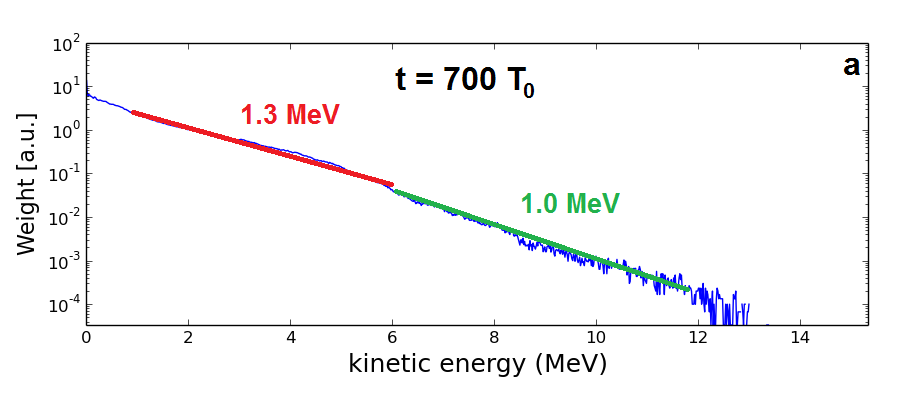}
\hspace{0.5cm}
\includegraphics[scale=0.32]{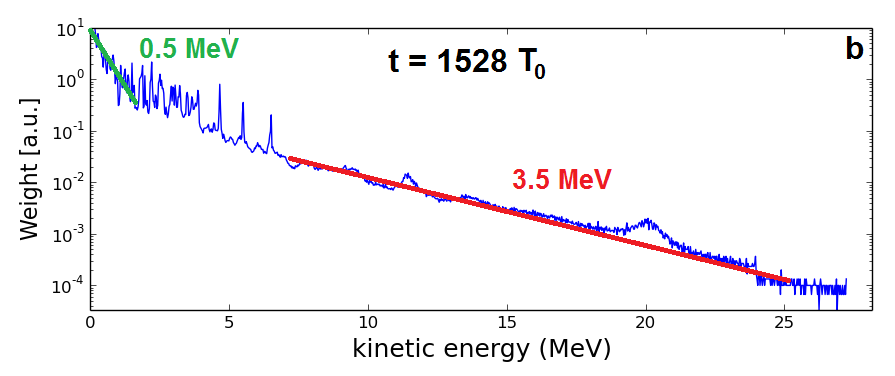}
\caption{Electron (a) and ion (b) distribution functions computed at $t \approx 700\, T_0$ and $t \approx 1528\, T_0$, respectively. The energy is expressed in the electron and ion relativistic factors, respectively.}
\label{fig_these-6_fctdistrib}
\end{figure}

Figure \ref{fig_these-6_fctdistrib}.a displays the electron distribution function at $t \approx 700\, T_0$, when the laser pulse goes out of the plasma. It shows that electrons have an approximately Maxwellian distribution with the temperature varying in the range between 1.0 and 1.3 MeV, while the maximum electron energy attains the level of more than 12 MeV. The average electron energy is consistent with the laser-plasma energy balance, confirming that the laser energy is deposited in the whole plasma volume. Indeed the total laser energy deposited in the plasma is about 1.9 MJ/cm$^2$, which corresponds to the average energy of 0.7 MeV per plasma particle. 

Apart of the most energetic electrons escaping from the plasma, the remaining electrons are recirculating and thus distributing energy evenly in the whole plasma volume. At the same time, they create a large charge separation field at the plasma edges, which accelerates protons according to the TNSA (Target Normal Sheath Acceleration) mechanism. This process, responsible for the energy transfer from electrons to ions, firstly occurs at the plasma front side before the laser pulse goes out of the plasma. Then, after $t \approx 800 T_0$, when the laser has left the plasma, the total electron energy starts to decrease (see Fig. \ref{fig_these-6_electronheating}.a) while the ion energy increases. However, the TNSA mechanism corresponds to acceleration of a relatively small number of ions which therefore can gain a large energy.

\begin{figure}
\centering
\includegraphics[scale=0.4]{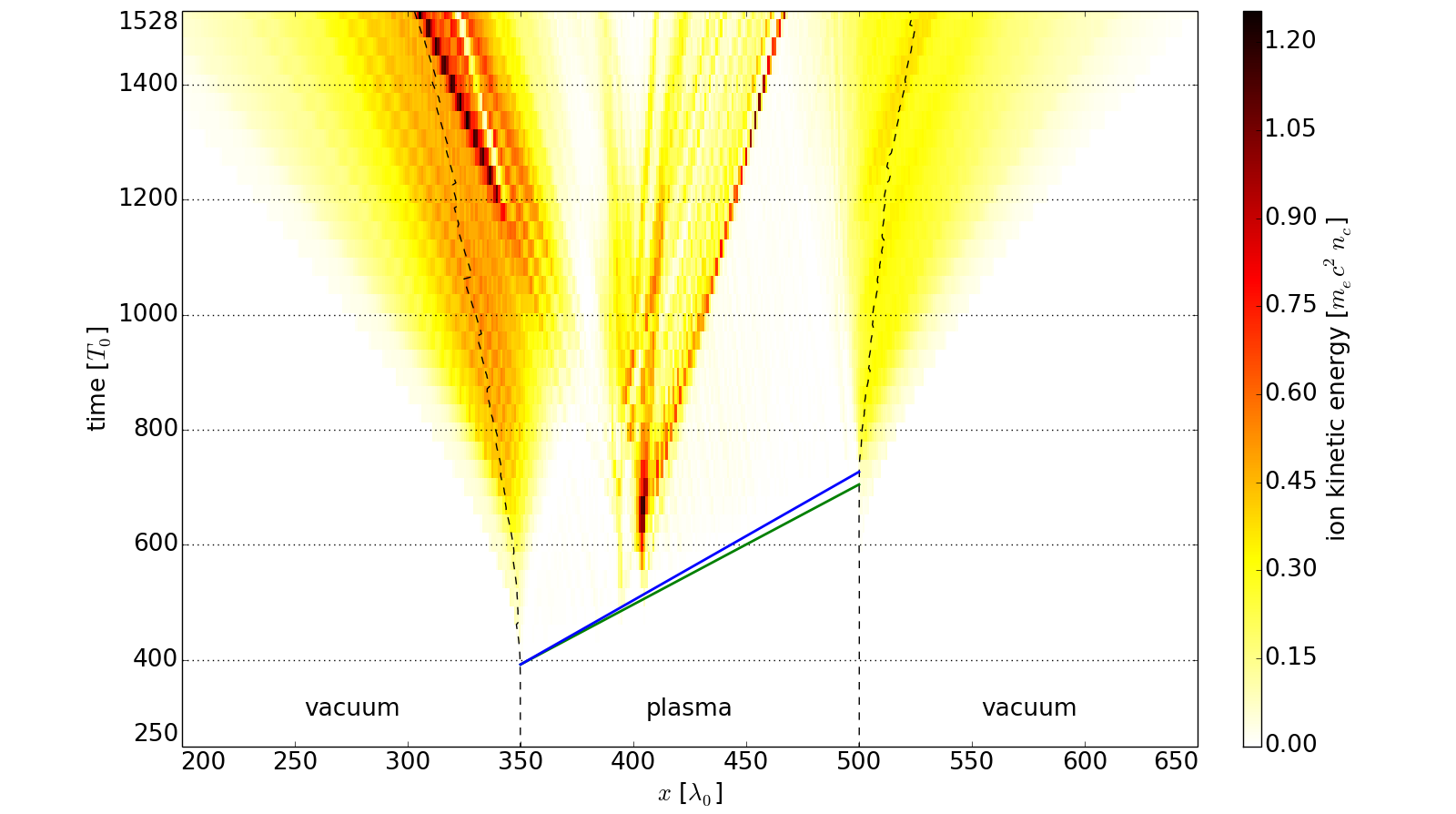}
\caption{Ion energy density as a function of space and time. Dashed lines delimit the plasma density $n_p$ where $n_p < 0.1\, n_c$. The green line (corresponding to the (2)-(3) boundary line in Fig. 1.a) shows the arrival of the high amplitude laser pulse. The blue line (corresponding to the (0)-(1) boundary line in Fig. 1.d) shows the triggering of the back-scattered electromagnetic waves by the plasma.}
\label{fig_these-6_ionheating}
\end{figure}

Figure \ref{fig_these-6_ionheating} shows the ion energy density as a function of space and time. The two bunches of ions accelerated by the charge separation field at each plasma edge are visible in yellow at $x<350\, \lambda_0$ and at $x>500\, \lambda_0$. Their energy cut-off reaches 27.2 (plasma front side) and 24.0 MeV (plasma rear side) at the end of the simulation. However, their relative number is small. The ion distribution function in Figure \ref{fig_these-6_fctdistrib}.b shows that the ion average energy is about 3.2 MeV in the energy range above 1 MeV. However, there are about 0.3 \% of protons with the energy exceeding 10 MeV. Moreover, there are several quasi-mono-energetic ion bunches with energies 5-7 MeV originating from the plasma cavities. This is confirmed in Fig. \ref{fig_these-6_ionheating}, which shows bunches of ions accelerated inside the plasma (visible in red), especially from the cavity positions. These processes of ion acceleration lead to an efficient transfer of the absorbed laser pulse energy to the protons. Their total energy reaches 29.5 \% of the total laser pulse energy (see Fig. \ref{fig_these-6_electronheating}.b), at the end of the simulation.

In conclusion, we have shown than the SRS instability is the main process responsible for the laser pulse absorption and energy transfer to electrons via the wave-breaking of the plasma waves. This electron energy is then efficiently transferred to ions due to the charge separation electrostatic field.

\section{Discussion and conclusion}
\label{sec_discussion}

Our analysis demonstrates the mechanism of an efficient energy transfer of an intense laser pulse to particles in a near-critical plasma. Although this mechanism has been demonstrated for a specific choice of laser and plasma parameters, its validity has been confirmed with other simulations carried out with different interaction parameters. They confirm the role of SRS as the major process responsible for the efficient laser energy absorption in a near critical plasma.

The simulation of the interaction of a circularly polarized laser pulse with the same intensity and duration, with the same plasma, gives very similar results. Electrons and ions get 35.4 \% and 28.3 \% of the laser pulse energy, at the end of the simulation, respectively, so that the absorption reaches 63.7 \%. The proton energy cut-off reach 31.8 MeV and 24.4 MeV at the front and rear plasma side, respectively.

\begin{figure}
\centering
\includegraphics[scale=0.4]{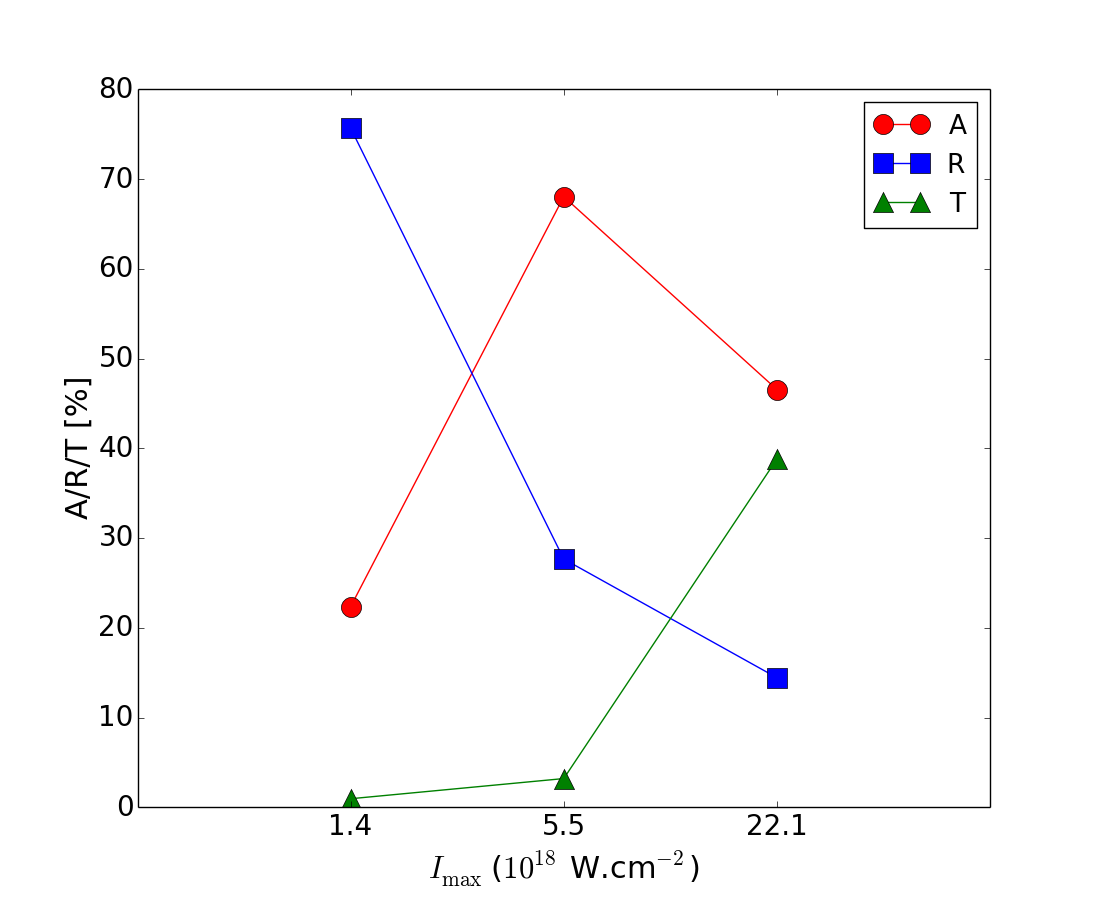}
\caption{Interaction energy balance as a function of the laser pulse intensity: $A$, $R$ and $T$ are respectively the plasma kinetic energy, the reflected and transmitted laser energy at the end of the simulation ($t_s \approx 1528\, T_0$). These quantities are normalised to the total laser pulse energy. The plasma density is $0.5\, n_c$ and the length is $150\, \lambda_0$.}
\label{fig_polarlin_comparison}
\end{figure}

Figure 10 shows the dependence of the absorbed and reflected laser energy on the maximum intensity for the 1 ps laser pulse duration, the plasma density $n_e/n_c=0.5$ and the plasma length of $150\, \lambda_0$. In the case where the maximum intensity of the laser pulse is multiplied by a factor of four ($I_{\max} \approx 2.2 \cdot 10^{19}$ W.cm$^{-2}$), we observe, all along the plasma, the same laser pulse depletion, electromagnetic back-scattering, and electron heating by the SRS instability, as shown in Figs. \ref{fig_these-6_Eyright-trans} and \ref{fig_these-6_electronheating}. However, a larger part of the laser pulse is transmitted through the plasma without being absorbed so that the cumulated transmission eventually reaches 38.8 \%. The total electron energy reaches around 42 \% of the total laser pulse energy when the laser pulse goes out of the plasma. This represents 2.8 times as much energy as the energy transferred to electrons in the $a_0=2$-case since the intensity and the energy of the laser pulse are multiplied by 4. The ion energy cut-off then reaches 64.7 MeV (plasma front side) and 83.9 MeV (plasma rear side) at the end of the simulation.

In the case where the laser intensity is divided by four ($I_{\max} \approx 1.4 \cdot 10^{18}$ W.cm$^{-2}$), the plasma absorption is reduced to 22.3 \%. A large part of the remaining laser pulse energy is backscattered by the plasma (see Fig. \ref{fig_polarlin_comparison}). In this case, the laser intensity is not sufficiently high to excite the SRS instability, the absorption is strongly reduced and the backscattered wave has a frequency close to the laser frequency $\omega_0$. The reduced laser energy deposition is readily manifested in the reduced ion energy cut-off which is 7.9 MeV and 5.0 MeV for the plasma front and rear side respectively.

The chosen plasma density of $0.5\, n_c$ is optimal for the efficient laser absorption, which is reduced for both lower and higher plasma densities. For the lower densities, absorption is still related to the SRS instability, but it is less efficient and more laser energy is transmitted through the plasma. In the case of interaction of a laser pulse with the intensity $I_{\max} \approx 5.5 \cdot 10^{18}$ W.cm$^{-2}$ with a denser plasma whose density equals $0.8\, n_c$, the absorption is reduced to 44.7 \%. Similarly as for the $0.5\, n_c$ case, the SRS instability is the main process responsible of the electron heating, however, it leads for a stronger back-scattering, which represents 53.7 \% of the total laser pulse energy. Besides, in this case, the frequency of the plasma wave excited by the SRS instability equals $0.3\,\omega_0$. Since this frequency is lower than the one excited in the $0.5\, n_c$ plasma ($0.5\, \omega_0$), the fraction of energy $\omega_p/\omega_0$ transferred to the plasma wave and then to the electrons is also lower than for the $0.5\, n_c$ case. The laser pulse is progressively depleted in the plasma as the zone of SRS activity extends with a lower velocity $\sim 0.3\, c$ so the laser pulse is almost totally absorbed along the first $100\, \lambda_0$ of the plasma length and the transmission equals only 0.7 \%. Consequently, the laser energy deposition is rather inhomogeneous leading to a less efficient ion acceleration. The protons reach a higher energy, 28.0 MeV, at the front side and a lower energy, 18.4 MeV, at the rear side than for the $0.5n_c$-case. This example confirms that the target areal density $n_{e0}l$ should be optimized for an efficient and homogeneous electron heating and an efficient ion acceleration.

These comparisons show that the particular case presented in this paper corresponds to the optimal choice of parameters, which leads to the highest energy transfer to the plasma. It corresponds to a case where the SRS instability leads to a very efficient and homogeneous electron heating and low rate of reflectivity (below 30 \%). These conditions are favourable for ion acceleration at the plasma edges and in the cavities. The present study is limited to 1D simulations, which allow us to have a high numerical precision along with a high spatial and temporal resolution. Thanks to large arrays of data, the features in real space can be compared with the detailed spectral properties of the fields thus allowing a clear identification of the physical mechanisms in play. In particular, this is the first clear demonstration of the dominant role of the SRS instability in plasmas with the density larger than the quarter of critical density.
We believe the physical processes discussed in this paper are also operational in the real three-dimensional space. Although the microscopic features may look different in 2D or 3D simulations due to other competing processes such as laser filamentation and plasma wave modulation, the characteristics averaged over the transverse directions should not be much different from the 1D simulations. This has been demonstrated for lower laser intensities for the case of SBS \citep{riconda2006} and SRS \cite{klimo2014} parametric instabilities.

\acknowledgments

This work was partly supported by the Project ANR-2011-BS04-014 from the French National Agency of Research and the Aquitaine Region Council. This work has been carried out within the framework of the EUROfusion Consortium and has received funding from the European Unions Horizon 2020 research and innovation programme under Grant Agreement No. 633053. We acknowledge the MCIA (M\'esocentre de Calcul Intensif Aquitain) of the Universities of Bordeaux and Pau et des Pays de l'Adour for providing computing facilities.

\end{document}